\documentclass[12pt,a4paper]{article}

\usepackage[textheight=23cm,textwidth=15cm]{geometry}
\usepackage{amsmath}
\usepackage{graphicx}
\usepackage[american]{babel}
\usepackage{here}
\usepackage[articletitle=true, super=false]{achemso}
\setcitestyle{square}
\usepackage{url}

\begin{document}

\begin{center}

{\LARGE\bf
A New Paradigm for\\ Computational Chemistry
}

\vspace{1cm}

{\large
Raphael T. Husistein$^{a,}$\footnote{ORCID: 0009-0005-2696-4574} and
Markus Reiher$^{a,}$\footnote{Corresponding author; e-mail: mreiher@ethz.ch; ORCID: 0000-0002-9508-1565}
}\\[4ex]

$^{a}$ Department of Chemistry and Applied Biosciences, ETH Zurich, \\
Vladimir-Prelog-Weg 2, 8093 Zurich, Switzerland

March 31, 2026

\vspace{.43cm}

\textbf{Abstract}
\end{center}
\vspace*{-.41cm}
{\small
Computational chemistry has become an indispensable tool for generating data and insights, pervading all branches of experimental chemistry. Its most central concept is the potential energy hypersurface, key to all chemistry and materials science, as it assigns an energy to a molecular structure, the necessary ingredient for reaction mechanism elucidation and reaction rate calculation. Density functional theory (DFT) has been the most important method in practice for obtaining such energies, which is mirrored in the use of high-performance computing hardware. In the last two decades, a new class of surrogate potential energy functions has been evolving with remarkable properties: quantum accuracy combined with force-field speed. Until very recently, their application was hampered by the fact that they needed to be trained on truly large system-specific data sets, generated before a computational chemistry study could be started (in sharp contrast to DFT, which, as a first-principles method, works out of the box, but at a far higher price of computational cost). Very recently, this roadblock has been overcome by so-called foundation machine learning interatomic potentials, which are poised to completely change the way we do computational chemistry, likely prompting us to abandon DFT as the prime method of choice for this purpose in less than a decade.
}

\newpage

\section{Introduction}
Central to our understanding of chemical processes is the assignment of a potential energy to a molecular structure, more specifically to a nuclear configuration, which leads to the concept of a potential energy surface that allows us to study reaction chemistry in terms of stable intermediates, reaction paths, transition state structures, and so forth. It is the Born-Oppenheimer approximation, the so-called clamped-nuclei approximation, which allows us to calculate the electronic energy of such a fixed arrangement of atomic nuclei. There exists a plethora of quantum chemical methods to calculate this electronic energy, all characterized by some balance of accuracy and feasibility. These methods allow us to predict reaction energies and activation barriers routinely \cite{Cramer2004, Jensen2007}, alongside molecular structures and transition states on the basis of the first principles of quantum mechanics, an accomplishment that was recognized by a share of the 1998 Nobel Prize in Chemistry to John Pople.

In the 1990s, approximate density functional theory (DFT) took the lead for this purpose within computational chemistry owing to its convincing compromise of bearable computational cost and sufficient reliability, leading to the recognition of its impact on chemistry by the other share of the 1998 Nobel Prize in Chemistry to Walter Kohn. 
More than 90\% of all quantum chemistry calculations are of DFT type \cite{QChem43manual}
and about one third of all computer science applications on supercomputer systems \cite{Camps2025}. Hence, the practical and economic importance as well as the ecological footprint of DFT can hardly be overestimated.

The accuracy (and to some degree the efficiency) of DFT depends on the approximate form chosen for the so-called exchange--correlation (xc) density functional, typically denoted by one of the notorious capital-letter acronyms. Although an exact xc functional can be shown to exist, its analytical form is not known and would probably also be too involved for efficient computational chemistry approaches. Hence, rather drastic and often heavily parametrized approximations are adopted in practice, while attempts are made to fulfill the
comparatively small number of known rigorous constraints. However, the approximate potentials, which are obtained from these approximate xc energy functionals as functional derivatives, differ significantly from reconstructed reference potentials \cite{Jacob2011, Boguslawski2013, Mori2014}. The fact that such xc functionals nevertheless produce useful results in chemical applications can be taken as an indication  
that there must be some sort of error compensation present. In fact, error compensation can be exposed if approximate exchange functionals are replaced by exact local exchange in Kohn--Sham DFT calculations, which then no longer compensates for deficiencies in the  approximation to the correlation functional \cite{Gorling1999}.

As a result, practical DFT calculations have a significant semi-empirical character.
An important consequence is that they cannot be systematically improved. However, it would already be sufficient to know the system-focused error of some specific DFT calculation \cite{Reiher2022}, but rigorous uncertainty quantification is not easy (see, for instance, Ref. \cite{Simm2016}) and Bayesian approaches have their limits \cite{Mardirossian2017, Weymuth2022}. Usually, one acquires empirical knowledge about the reliability of DFT results over time, by which one can establish some 'gut feeling' regarding the reliability of certain xc functionals in specific calculations. We emphasize this touch of an anthopogenic bias to prepare the ground for a higher acceptance of the paradigm shift brought about by machine-learning inference models.

The last two decades have seen an intense development of what is now called machine learning interatomic potentials (MLIPs) \cite{Unke2021, Behler2021}. 
In their seminal paper from 2007, Behler and Parrinello \cite{Behler2007}
proposed high-dimensional neural network potentials that were applicable to large atomistic systems (by virtue of an architecture that scales linearly in the number of atoms), yielded accurate results on par with DFT (which was used as reference for their system-focused training), and obeyed the proper symmetries.
The value of their approach became clear quickly, because these neural network potentials were able to deliver the accuracy of quantum chemical methods combined with the computational efficiency of classical force fields. However, they came with a huge burden regarding their parameterization. Thousands of quantum calculations were needed to parametrize such MLIPs, which created a huge barrier for general purpose applications --- but this situation has changed in the past few years. Foundation models were developed which demonstrate remarkable extrapolation capabilities, so that initial pre-training is no longer mandatory \cite{Batatia2025b, Wood2025}. Even chemical reactions can now be studied out of the box. 
The tremendous success of these approaches make them the natural enemy of the current canonical choice for computational chemistry applications, that is DFT. 

In this work, we provide a brief review of MLIPs for the approximation of potential energy surfaces. We discuss their strengths and current weaknesses in order to eventually paint a picture of future computational chemistry relying on foundation MLIPs as the entry point for calculations in molecules and materials.

\section{Machine Learning Interatomic Potentials}

We first review and discuss some basic theoretical foundations of the new powerful class of potential energy functions.
Various names have been used in the literature that all refer to the same concept: machine learning potential (MLP), machine learning force field (MLFF), and machine learning interatomic potential (MLIP). Since MLP could be mistaken for ''multi-layer perceptron,`` there has been a shift towards the term machine learning interatomic potential and the corresponding abbreviation ''MLIP.`` MLIPs can be developed using various machine learning algorithms, such as linear regression, Gaussian process regression, and neural networks. Recently, however, MLIPs based on neural networks have become predominant in the field, so we will focus on those.

MLIPs must meet two requirements in order to describe atomic interactions effectively and accurately. First, they need to encode a molecule or atomistic assembly in a way that enables machine learning algorithms that rely on fixed-size inputs to work with systems of arbitrary size. Second, MLIPs must address the physical symmetries inherent to the problem. Specifically, the predictions of the energy should remain invariant under translations and rotations of the entire atomistic system and when identical atoms are swapped. The first requirement is typically met by expressing the total energy as a sum of local (atomic) contributions,
\begin{equation}
    E_{\text{tot}} = \sum_{i=1}^{N} E_i\!\left(\{\mathbf{r}_j \mid \|\mathbf{r}_j - \mathbf{r}_i\| < r_c\}\right),
    \label{eq:energy-decomposition}
\end{equation}
where each atomic contribution $E_i$ depends only on atoms within a cutoff radius $r_c$ around atom $i$. 
Physically, this approach can be motivated by the ``nearsightedness of electronic matter'' \cite{Prodan2005}. However, this assumption of locality implies that long-range interactions are neglected. While semi-local DFT functionals similarly miss long-range correlations such as dispersion \cite{Grimme2016}, purely local MLIPs also omit long-range electrostatics.
For the second requirement, there is a wider array of techniques. For example, early work in the mid-1990s and early 2000s employed carefully chosen coordinate systems that respect the system's symmetries \cite{Blank1995, Brown1996, Lorenz2004}. 

A shortcoming of this approach has been that it could not be easily applied to different systems. This limitation has been mitigated by the introduction of descriptors, which are representations of the local atomic environment that respect the necessary symmetries \cite{Behler2007}. One prominent class of such descriptors is the so-called atom-centered symmetry functions (ACSF) \cite{Behler2007, Behler2011}. Although many different types of descriptors exist, including ACSF-based descriptors \cite{Behler2007, Behler2011, Smith2017, Gastegger2018, Eckhoff2023},  the smooth overlap of atomic positions (SOAP) \cite{Bartok2013}, the spectral neighbor analysis potential (SNAP), moment tensor potentials \cite{Shapeev2016}, and the atomic cluster expansion (ACE) \cite{Drautz2019}, they share the common goal of encoding the local atomic environment in a symmetry-respecting way. 

To make this concrete, consider the high-dimensional neural network potential (HDNNP) of Behler and Parrinello \cite{Behler2007}, in which each atomic energy $E_i$ (Eq.~\eqref{eq:energy-decomposition}) is predicted by a separate neural network NN taking a descriptor vector $\mathbf{G}_i$ as input,
\begin{equation}
    E_i = \text{NN}(\mathbf{G}_i;\, \boldsymbol{\theta}), \qquad
    \mathbf{G}_i = \bigl[G_i^{(1)},\, G_i^{(2)},\, \ldots\bigr],
    \label{eq:hdnnp}
\end{equation}
where $\text{NN}(\cdot\,; \boldsymbol{\theta})$ denotes a neural network with learnable 
parameters $\boldsymbol{\theta}$. A typical radial symmetry function that encodes the neighbor distribution around the atom $i$ is
\begin{equation}
    G_i^{(1)} = \sum_{j \neq i} e^{-\eta(\|\mathbf{r}_{ij}\| - \mu)^2}\, f_c(\|\mathbf{r}_{ij}\|),
    \label{eq:symm-func}
\end{equation}
where $\|\mathbf{r}_{ij}\|$ is the distance between the atoms $i$ and $j$, $\eta$ and $\mu$ are hyperparameters controlling the width and center of the Gaussian, and $f_c$ is a smooth cutoff function making sure that contributions vanish continuously at $r_c$. 
To capture angular information, the radial functions are supplemented with angular functions:
\begin{equation}
    G_i^{(2)} = 2^{1-\zeta}\sum_{j,k \neq i} (1 + \lambda\cos(\theta_{ijk} ))^\zeta e^{-\eta(\|\mathbf{r}_{ij}\|^2 + \|\mathbf{r}_{ik}\|^2 + \|\mathbf{r}_{jk}\|^2})\, f_c(\|\mathbf{r}_{ij}\|)f_c(\|\mathbf{r}_{ik}\|)f_c(\|\mathbf{r}_{jk}\|),
    \label{eq:symm-func-angular}
\end{equation}
where $\lambda \in \{-1, 1\}$, $\eta$ and $\zeta$ are hyperparameters that control the spatial and angular sensitivity of the function to the local atomic environment, and $f_c$ is again a smooth cutoff function.
By 
construction, $\mathbf{G}_i$ is invariant under translation, rotation, and permutation of same-element atoms, so Eq.~\eqref{eq:hdnnp} satisfies all required symmetries.

\begin{figure}
    \centering
    \includegraphics[width=\linewidth]{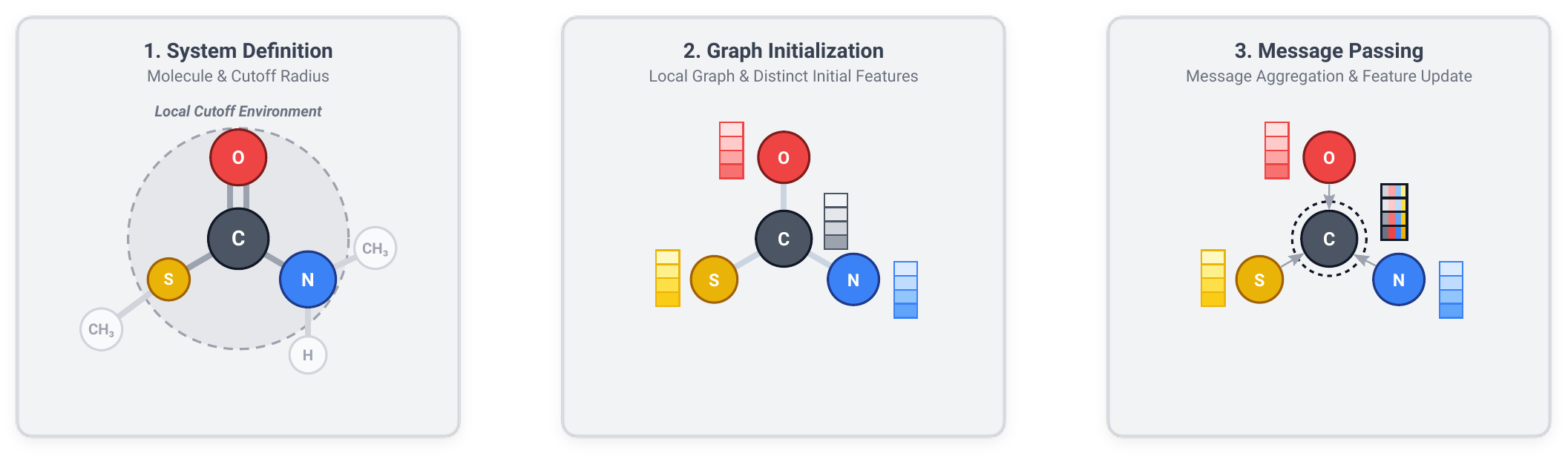}
    \caption{Illustration of the message passing scheme applied to a molecule. (1) A local environment is defined for each atom using a cutoff radius. (2) The resulting graph is constructed, and each atom is assigned an initial feature vector. (3) Each atom passes its feature vector to its neighbors. An atom then uses its own feature vector and those gathered from its neighbors to update its representation. For clarity, only the update of the carbon atom is shown.}
    \label{fig:message-passing}
\end{figure}

Instead of using manually designed descriptors, one may learn descriptions of the local environment from available training data. In this approach, the system is encoded as a graph, with atoms as nodes and edges connecting all nodes within a cutoff distance. This graph is then used as input for a graph neural network \cite{Scarselli2008}. Each atom is initially assigned a feature vector, either derived from atomic properties or randomly initialized, that serves as a numerical representation of the atom within the system.
During training, atoms connected by an edge exchange their feature vectors, a process usually called ``message passing,'' and then update their own features based on the received features \cite{Duvenaud2015, Gilmer2017} (see Figure \ref{fig:message-passing} for an illustration of this process). 

Initially, graph-based models used features, such as distances and angles, that were invariant with respect to the symmetries of the systems \cite{Schutt2018, Unke2019, Chen2022}. Since energy remains invariant under rotation, translation, and permutation of identical atoms, 
this approach does not present challenges for energy predictions. However, tensorial properties, such as forces, change orientation under rotation of the system, making predictions based on invariant features impossible. Although models that use invariant features avoid this issue by calculating the force experienced by an atom as the negative gradient of energy with respect to the atom's position, they cannot directly predict other tensorial properties, such as dipole moments or the polarizability tensor.

This limitation is addressed by equivariant models, which are models whose features transform in the same way as the input \cite{Thomas2018, Anderson2019}. Many of today's state-of-the-art equivariant models are constructed using features based on spherical tensors represented by spherical harmonics \cite{Batzner2022, Batatia2022, Musaelian2023, Liao2024, Park2024, Wood2025, Batatia2025}. To maintain equivariance when aggregating tensorial features, the Clebsch-Gordan tensor product is used, but computing it, which is required during training and inference, is computationally expensive. An alternative line of work uses Cartesian tensors instead, which is conceptually simpler and computationally less demanding. However, most architectures are restricted to tensors of rank 1 \cite{Schutt2021, Satorras2021, Tholke2022} or rank 2 \cite{Simeon2023}, and extensions to higher-order tensors are an exception \cite{Zaverkin2024, Wang2024, Cheng2024}. 

Empirically, equivariant models have been found to improve data efficiency, which means that fewer training samples are required to achieve the same level of performance \cite{Batzner2022}. Although enforcing symmetries through the input features or the architecture remains the standard approach, there are alternatives. For example, penalty terms can be added to the training procedure, discouraging violations of symmetry during optimization and thus guiding the model to learn the necessary symmetries \cite{Elhag2025}. Another option is to use data augmentation, similar to what was done in AlphaFold 3, which also allows the model to learn symmetries from data \cite{Abramson2024, Mazitov2025}.

\section{The Breakthrough: Foundation MLIPs}

The term ``foundation model'' has its origins in the field of natural language processing, where large models such as BERT were trained on large amounts of text and then achieved state-of-the-art results on a wide variety of tasks, such as question answering and natural language inference (i.e., determining logical relationships between sentences, such as whether one statement entails or contradicts another), through fine-tuning \cite{Bommasani2021, Devlin2019}. This approach marked a paradigm shift: models are trained on large and diverse datasets and then fine-tuned for specific tasks. Scaling up both the model and the training data resulted in emergent capabilities for which the models were not explicitly trained \cite{Bommasani2021}, which greatly contribute to their success and their increasing popularity.

In the realm of MLIPs, the initial focus was on training models for specific chemical systems, limited to certain elements and geometrical arrangements. However, with the advent of foundation MLIPs (sometimes also called universal MLIPs), there has been a shift toward applying large, general models that do not require extensive system-specific training. The first generation of foundation MLIPs demonstrated that general-purpose interatomic potentials are achievable in principle. The Materials Graph Network (MEGNet) was trained on data from approximately $60\, 000$ crystals in the Materials Project database, covering $89$ elements of the periodic table \cite{Jain2013, Chi2019}. It showed that a single model could deliver low prediction errors on a wide range of properties. Since then, a plethora of new foundation MLIPs trained on large datasets have been published, including M3GNet, MACE-MP-0, and eqV2, for example \cite{Chen2022, Batatia2025b, Barroso2024}. 

These models have then advanced in several ways: they have grown in parameter size, adopted newer architectures, and trained on larger datasets. While the original MEGNet model had approximately 161,000 trainable parameters, M3GNet had 228,000; MACE-MP-0 had 4.7 million; and eqV2-L had about 153.8 million. Similar trends can be observed in the sizes of the training sets used. These advances have brought about a paradigm shift in the field, enabling stable molecular dynamics, high-throughput materials discovery, and property predictions across a vast chemical space, all without the need for retraining \cite{Chen2022}. Furthermore, MACE-MP-0, trained primarily on the MPtrj dataset with a strong materials focus, showed encouraging out-of-distribution transferability, including qualitatively stable MD of liquid water \cite{Deng2023, Batatia2025b}. 

Although these models deliver good results on many systems, the training data used ultimately place constraints on what they can be used for. For example, systematic underestimation of energies and forces was observed in M3GNET and MACE-MP-0, and since the models were largely trained on DFT data from bulk materials, they have difficulty predicting surface energies \cite{Deng2025, Focassio2025}. Another study found that predicting the adsorption energies of different N$_x$H$_y$ species on Ni$(111)$, Ru$(0001)$, Ru$(10\overline{1}0)$ and Ru$(10\overline{1}5)$ requires additional fine-tuning for all tested models, including eqV2 \cite{Zills2025}.

Further progress in addressing the limitations imposed by the training data used came with the Universal Model for Atoms (UMA), a model trained on a dataset of half a billion structures \cite{Wood2025}. Not only is this data set of previously unattained size, but it also spans a wide range of domain-specific areas, including materials, molecules, catalysis, molecular crystals, and metal-organic frameworks. For this purpose, the five datasets OMat24, OMol25, OC20, OMC25, and ODAC25 were combined, and a single model was trained on all of them \cite{Barroso2024, Levine2025, Chanussot2021, Chanussot2021b, Gharakhanyan2025, Sriram2025}. While training is usually performed with data generated using a single exchange-correlation density functional as a DFT reference, here, training was performed on a combined dataset generated using different density functionals. Analysis showed that knowledge transfer occurred across the datasets, with a model trained on all datasets outperforming one trained exclusively on molecular crystal data \cite{Wood2025}. While there are three UMA model variants, ranging from 150 million to 1.4 billion parameters, even the smallest variant was shown to achieve results generally comparable to, or better than, specialized models across a broad range of domains without any fine-tuning \cite{Wood2025}. This has since been corroborated. For instance, the smallest UMA model variant was applied to transition-metal complexes without undergoing additional training. The results showed that reliable conformer ranking is possible for most ligands in rigid systems but is less reliable for flexible or fluxional systems, where energy differences fall below a few kJ/mol and thus approach the accuracy of DFT itself \cite{Kalikadien2026}.

This development constitutes a paradigm shift within the field of MLIPs, demonstrating that large, general models can be applied throughout many domains, much like foundation models in natural language processing. Several such models have now been released, each presenting different trade-offs in accuracy and speed \cite{Deng2023, Yang2024, Barroso2024, Park2024, Batatia2025, Rhodes2025, Fu2025, Wood2025, Koker2025, Mazitov2025, Zhang2025, Lysogorskiy2026, Qu2026}. Early applications demonstrate the versatility of these models. For example, MACE-MP has been shown, without system-specific retraining, to accurately simulate the metastability of siliceous zeolites relative to $\alpha$-quartz and the phase transition pressures of silica polymorphs under high pressure, with results aligning well with experimental values \cite{Nasir2025}. Similarly, a systematic comparison of approximately 10,000 semiconductors indicates that MatterSim-v1 accurately predicts harmonic phonon properties, with errors smaller than the difference between the exchange-correlation functionals PBE and PBEsol, suggesting that some foundation models have achieved DFT-level fidelity in vibrational properties \cite{Loew2025}.

\subsection{Static Generalist Models vs. (Repeated) System-Focused Fine Tuning of Foundation MLIPs}

As discussed in the previous section, foundation models have arrived in chemistry, and they are here to stay and make a long-lasting impact. However, the question now arises as to how these models should be applied. Should they be used as static, generally applicable models, or should they be fine-tuned to specific systems?  We structure the comparison along four decision axes that directly influence this choice: (1) accuracy relative to application tolerances, (2) maintenance risk due to catastrophic forgetting under iterative updates, (3) data and computing constraints, and (4) the need for reliable uncertainty estimates.

\subsubsection{Accuracy}
As mentioned earlier, foundation models often achieve an accuracy level that is equivalent to or better than that of system-specific models. However, ultimately, it is important to assess accuracy on a case-by-case basis against application tolerances. For instance, the accuracy of predictions for diatomic molecules outside equilibrium bonding has been found to be inaccurate for some foundation models\cite{Lee2025}. Although some models have addressed this issue by including diatomic data in the training set \cite{Mazitov2025, Wood2026}, this illustrates that accuracy must be assessed critically against application-specific needs and tolerances. However, if a static generalist model meets the target error thresholds in the relevant chemical areas, using a generalist model can considerably reduce the required effort.

\subsubsection{Catastrophic Forgetting}
If a generalist model cannot achieve the desired accuracy, an alternative is to fine-tune it to incorporate new knowledge. However, this increases the risk of ``catastrophic forgetting'', i.e., a deterioration in performance in areas previously mastered \cite{McCloskey1989}. If the model is to remain generally applicable yet be expanded to include specific knowledge, catastrophic forgetting must be prevented while simultaneously keeping it flexible enough to incorporate new information. This tension is known as the stability-plasticity dilemma \cite{Grossberg1980}, 
addressed in the field of continual learning, for which many strategies have been proposed in the literature. 

These strategies can be categorized as (i) regularization-based, (ii) replay-based, and (iii) parameter isolation \cite{De2021}. Regularization-based approaches add penalties or constraints to the loss function (the objective that is minimized during training) to prevent modification of important weights associated with learned knowledge \cite{Kirkpatrick2017, Zenke2017}, or to constrain the model's predictions on previously learned data \cite{Li2017}. Replay-based approaches periodically retrain on past training data to prevent forgetting \cite{David2019, Shin2017}. Parameter isolation approaches dedicate separate subsets of model parameters to each task, preventing interference between old and new knowledge \cite{Rusu2016, Yoon2018}. 

Most continual learning strategies have been developed for traditional machine learning tasks, such as image classification, and have only recently been applied to MLIPs \cite{Eckhoff2023, Kim2025}. The most straightforward approach, and the one most frequently encountered in the MLIP literature, is to freeze large portions of the network and only adjust the weights in the final layers \cite{Deng2024, Radova2025, Bensberg2025a}. Although this prevents catastrophic forgetting in the frozen layers, it significantly limits the ability to incorporate new knowledge. 

A more targeted alternative strategy, called low-rank adaptation (LoRA), has been developed for large language models  \cite{Hu2022}. Instead of adjusting the weights directly, they are frozen, and small, trainable, low-rank matrices are introduced. This enables structured, low-rank updates across all layers while preserving the pretrained weights. This approach provides a more flexible balance between stability and plasticity than layer freezing. Although applying LoRA directly to equivariant MLIPs would break their equivariance, recent work has extended the concept to work with equivariant MLIPs. \cite{ChenWang2025}. However, the application of continual learning methods to foundation MLIPs is still in its early stages, and their ability to balance stability and plasticity in practice has yet to be systematically evaluated.

\subsubsection{Data and Compute Cost}
Static generalist models are ideal for large systems where generating training data would be computationally prohibitive. However, compared to system-specific models, training time is traded for longer inference times as the larger model size results in slower prediction speeds. In applications such as molecular dynamics, where chemical diversity is limited and inference time is critical, knowledge distillation is a promising approach. The goal is to transfer knowledge from a teacher (a static generalist) to a student model with fewer parameters. 
In doing so, general applicability is traded for significantly faster inference while maintaining, or in some cases exceeding, the teacher's accuracy within the target chemical domain \cite{Hinton2015, Ekstrom2023, Amin2025}. This points to an emerging paradigm for MLIP development, in which published foundation models are used to create smaller, system-specific MLIPs that enable fast sampling without the need for additional training data.

\subsubsection{Uncertainty Quantification}
\label{subsubsec:uncertainty-quantification}
Having an uncertainty estimate for predictions is important for many applications. For example, many active learning workflows use uncertainty to add new structures to the training data set \cite{Reker2015, Gastegger2017, Podryabinkin2017, Smith2018, Bensberg2025b}. The most common approach is to use an ensemble of models, either trained on different data splits or with different initializations, and use the empirical variance calculated from individual predictions \cite{Behler2014, Lakshminarayanan2017, Peterson2017}. 

However, in practice, a single static generalist model is typically released, which makes ensembling not applicable in this context. Although there are other techniques, such as Monte Carlo dropout \cite{Gal2016, Wen2020}, mean-variance estimation \cite{Nix1994}, and loss trajectory analysis for uncertainty measures \cite{Vita2025}, these generally need to be integrated during model training. By contrast, system-oriented fine-tuning allows for the inclusion of additional (lightweight) prediction heads and the subsequent use of ensemble methods \cite{Kellner2024, Bilbrey2025}. However, the quality of these uncertainties must be carefully evaluated. For example, it has been reported that uncertainties can be significantly underestimated \cite{Bilbrey2025}. When uncertainties must be estimated for a single static generalist without fine-tuning, several post-hoc approaches are available: For example, gradient information can be used to measure the sensitivity of predictions as an approximation of uncertainty \cite {Zaverkin2021, Bigi2024}. Alternatively, distance-based methods flag structures as uncertain if they are far from the training data in the model’s learned feature space. These methods do not modify the model and only require a set of reference calculations for calibration \cite{Janet2019, Yuge2022}. While gradient-based methods only need to iterate once over the training data to precompute some quantities, not at inference time, distance-based methods always need access to the training data. However, even a single iteration through the training data can be computationally demanding if the data comprises millions of structures.

Gradient-based uncertainty methods are the natural first choice for a static generalist model because they do not require model training or access to training data during inference. However, there must be sufficient computational power to iterate through the entire training dataset once. To validate the uncertainty estimates and assess the accuracy of the generalist model for the target system, a small set of reference calculations is nevertheless recommended. If the model's predictions are not accurate enough, fine-tuning can be attempted using the reference calculations generated as training data.

\subsection{Performance of MLIPs}
Comparing the computational cost of \textit{ab initio} methods and classical force fields with that of MLIPs is not as straightforward as one might think. First, these conceptually different approaches have different hardware requirements. For instance, most DFT computer programs run efficiently on central processing units (CPUs) only, whereas MLIPs perform best on graphical processing units (GPUs).  Second, it is not clear which measure is most suitable for the comparison: Should the focus be on runtime or on the computing power measured, for instance, in floating-point operations per second? What about power consumption, which can vary greatly between CPUs and GPUs? In addition, what degree of accuracy and system size should be compared? Since these methods have different time complexities, changes in system size can drastically alter the results. Given these difficulties and the fact that exact runtime differences depend on the specific application, our goal here is not to provide exact numbers. Instead, we intend to provide guiding estimates.

To understand how MLIPs perform compared to classical force fields and DFT, we performed structure optimizations on 101 naphthalene structures from the MD17 dataset \cite{Chmiela2017}. For optimizing structures with MLIPs, we used the Atomic Simulation Environment (ASE) program with the BFGS optimizer \cite{Larsen2017}. For DFT optimizations, we used BFGS and the NWChem interface of ASE with the PBE0 exchange-correlation functional 
and the def2-SVP dataset \cite{Valiev2010, Adamo1999, Weigend2005}. To allow for comparisons with classical force fields, we optimized the same structures using the universal force field (UFF) and the conjugate gradient algorithm in Open Babel \cite{Rappe1992, OBoyle2011}. 

The measured timings are shown in Table \ref{tab:mlip-ff-timings}. The overall conclusion is that, while force fields can be more than 200 times faster than the MLIPs studied, DFT calculations can be up to a factor of 1000 slower than MLIP calculations.

When we compare the time required for a single optimization step with different MLIPs, significant differences in performance become apparent between the various models. For example, an optimization step using the CPU with eqV2-M takes more than ten times longer than with Nequix. However, this is to be expected, since Nequix has around $700$ thousand parameters, while eqV2-M has $86$ million. It is also evident that the models do not all benefit equally from using a GPU (right column in Table \ref{tab:mlip-ff-timings}). On a GPU (compared to the CPU), GRACE-2L-OAM is more than five times faster, while Nequix is only about $20\,\%$ faster. However, on average, the MLIPs are more than three times faster on a GPU compared to the CPU (geometric mean of the speedup ratios: $3.07\times$). 

It is important to note that the provided timings are not intended for direct comparison between MLIPs, as they differ in parameter counts and accuracy levels. Moreover, the timing is affected by system size, leading to different relative timings in larger systems. Our main objective here is to provide a general idea of how MLIPs compare to force fields and DFT.

Our comparison of MLIP timings with those of the UFF further supports the existing literature that indicates that classical force fields are $100$ to $1\,000$ times faster \cite{Wang2025}. Similar substantial time differences can be observed between DFT and MLIP results, with the MLIP being orders of magnitude faster.

\begin{table}[]
    \caption{Comparison of runtimes in milliseconds of a single optimization step for naphthalene structures from the MD17 dataset on an Intel Core i7-14700K processor (middle column) and an NVIDIA RTX 2070 graphics card (right column). MLIPs (upper part) used the BFGS optimizer, while the UFF (second row from the bottom) used the conjugate gradient optimizer.
    \label{tab:mlip-ff-timings}}
    \centering
    \begin{tabular}{l|r|r}\hline\hline
    Method    & 4 Cores & GPU   \\ \hline
    MACE-MH-1  \cite{Batatia2025}                     & 63.79 & 20.14 \\
    GRACE-2L-OAM \cite{Lysogorskiy2026}    & 52.57   & 9.17  \\
    MatterSim v1 5M \cite{Yang2024} & 32.83   & 16.57 \\
    eqV2 M \cite{Barroso2024}       & 471.18  & 76.56 \\
    Orb-v3 \cite{Rhodes2025}        & 48.74   & 10.05 \\
    SevenNet-MF-ompa \cite{Kim2024} & 134.96  & 69.24 \\
    UMA-S-1p2 \cite{Wood2025, Wood2026}      & 118.07  & 30.63 \\
    Nequix \cite{Koker2025}        & 43.55   & 35.95 \\
    PET-MAD-1.5-S \cite{Mazitov2025}   & 28.78  & 11.63 \\\hline
    UFF \cite{Rappe1992}           & 0.15    & ---   \\\hline 
    PBE0 \cite{Adamo1999}          & 29301.416 & --- \\ \hline\hline
    \end{tabular}
\end{table}

\section{Challenges pointing to Future Developments}

Despite their impressive performance, foundation models for atomistic simulations still face several unresolved challenges. These challenges include limitations in the description of systems, computational scalability, data quality, and reliability. Overcoming these challenges will be crucial to realizing the full potential of MLIPs in real-world applications.

\subsection{Physical Fidelity}
A significant issue is their dependency on a fixed cutoff radius, typically ranging from 5 to 18 angstroms. Although message passing enables models to use information beyond this cutoff, information diffusion is ultimately limited by the cutoff and the number of message passing steps. This restricts the inclusion of long-range electrostatics and dispersion interactions, which are crucial, however, in the simulation of large atomistic assemblies such as metal-organic frameworks and biomacromolecules \cite{Formalik2018, Zhou2018}. 

Various studies have addressed this issue, for example, by augmenting MLIPs with analytical long-range dispersion terms \cite{Wen2019, Deringer2020}, by predicting partial charges or Wannier centres and summing
the resulting electrostatic interactions \cite{Artrith2011, Unke2019, Zhang2022} or by building nonlocal descriptors that can target both \cite{Grisafi2019, Huguenin2023}. An alternative to explicit charge prediction is the Latent Ewald Summation (LES) method, which instead infers latent charges purely from total energies and forces \cite{Cheng2025}. This removes the need for explicit charge labels in the training data, making LES compatible with the majority of published datasets. Other approaches use charge equilibration (Qeq) schemes \cite{Rappe1991}, in which electronegativities and hardness are predicted from local geometry \cite{Ghasemi2015, Ko2021, Staacke2022} and then used to minimize a quadratic energy functional with respect to atomic charges subject to the constraint of a fixed total charge. This enables the model to capture long-range charge transfer and redistribution across the system, but increases the computational overhead. 

Although most foundation models currently available do not directly account for long-range effects, the recently published models MACE-POLAR-1 and AllScAIP already do so, using two contrasting strategies \cite{Batatia2026, Qu2026}. MACE-POLAR-1 uses a physics-driven approach that extends the MACE architecture by including explicit long-range interactions. Atomic multipoles are iteratively updated by the local electric field using a non-self-consistent field formalism. This is followed by global charge and spin equilibration, which is enforced via learnable Fukui functions \cite{Batatia2026, Baldwin2026}. AllScAIP, on the other hand, relies on a data-driven strategy. It combines a local neighborhood attention step with a distance-based cutoff and an all-to-all node attention step with no cutoff. This enables the model to learn long-range interactions from data \cite{Qu2026}. It is important to note that incorporating information beyond a local cutoff means the computational complexity of the model no longer scales strictly linearly with the number of atoms, ranging from $\mathcal{O}(N \log N)$ for Ewald-based methods up to $\mathcal{O}(N^2)$ for AllScAIP, at least as long as no other measures, such as fast multipole approaches, are being taken. 
Regarding future developments, the jury is still out on the best approach to incorporating long-range effects into foundation models. Should the known physical form of long-range interactions be encoded explicitly, or should sufficiently flexible architectures be used to learn the functional form of long-range interactions directly from data? The latter might have advantages, as it could capture such effects in a very general sense (not bound to, and therefore beyond, analytic dispersion expressions), but accurate reference data for training will be difficult to generate.

Another limitation is the neglect of the total charge and spin multiplicity in many MLIPs \cite{Batatia2025b, Lysogorskiy2026, Yang2024, Barroso2024, Kim2024, Koker2025}. While incorporating these parameters as additional input values is straightforward at the architectural level, as some models do \cite{Wood2025, Rhodes2025, Batatia2025}, spin multiplicity alone is insufficient for many magnetic phenomena. Simulating antiferromagnetism, for example, requires explicit information about local magnetic moments on individual atoms. Several works have addressed the construction of magnetic MLIPs that incorporate local spin degrees of freedom \cite{Eckhoff2021, Novikov2022, Yu2022, Yu2024, Xu2025}, although these are system-specific models rather than foundation models. A first step in this direction was taken with the CHGNet model \cite{Deng2023}, which predicts collinear magnetic moments. However, the magnetic moments are not input parameters in this model, and therefore, the model cannot distinguish between structures with the same geometry but different spin configurations. Although recent foundation models have advanced in incorporating total charge and spin as inputs, incorporating local magnetic moments for solid-state magnetic simulations and extrapolating these models to unseen spin configurations and charge states remains a challenge.

\subsection{Scalability and Efficiency}
MLIPs are usually trained using energy and forces. However, many applications, such as transition state (TS) searches, require Hessians. For models in which the force is predicted rather than calculated as a gradient of the energy, incorporating Hessian information during training has been shown to be crucial for successful TS searches \cite{Cui2025}. Even for models in which the force is computed as the gradient of the energy, incorporating Hessian information improves the performance of TS searches \cite{Cui2025, Rodriguez2025}. Furthermore, incorporating Hessians into the training process has also been shown to improve vibrational frequency predictions \cite{Williams2025}. However, calculating the Hessian of a large MLIP scales quadratically with respect to the number of atoms and requires significant amounts of memory. Hence, training large foundation models using Hessian data and efficiently obtaining Hessians during inference remain challenges to be solved in the future.

Although comparisons of MLIPs are often limited to energy and force errors, it has been shown that models with small force errors can still fail in molecular dynamics simulations \cite{Fu2023}. Therefore, benchmarks that more closely resemble actual applications are essential. One notable benchmark that contributes to this effort is the MLIP Arena \cite{Chiang2025}. Looking ahead, it will be important to develop more advanced training paradigms that not only utilize energies and forces but also adapt models directly to specific tasks, such as molecular dynamics simulations.

\subsection{Data and Evaluation}
The accuracy of MLIPs depends on the training dataset. The current approach to building foundation models is to train them on increasingly large datasets that span a wide variety of chemical systems. However, this approach presents several challenges. First, training on ever-increasing amounts of data is computationally expensive. Without a clearer understanding of which data is important for training, a lot of computing power could potentially be wasted. Second, as datasets grow, the likelihood that erroneous or noisy data are included could increase, which ultimately limits the ability of MLIPs to learn. For instance, a recent study revealed that several datasets contained structures with non-zero net forces, suggesting numerical noise in the DFT calculation and suboptimal DFT settings \cite{Kuryla2025}. Therefore, future work must develop a strategy for constructing minimal datasets to reduce training time. At the same time, there needs to be a fully automated mechanism to ensure that the data are of sufficient quality.

Current foundational models largely lack an integrated mechanism for quantifying uncertainty. Although there are numerous methods for measuring prediction uncertainty (see Section \ref{subsubsec:uncertainty-quantification}), most models are published without an integrated uncertainty estimation method. As a consequence, one must resort to post-hoc approaches, such as readout ensembling (\cite{Bilbrey2025}). PET-MAD is a notable exception as a foundation model that has a directly integrated uncertainty estimation \cite{Mazitov2025}. However, foundation models must be equipped with well-calibrated uncertainty estimates to be practical to use. At the same time, developing uncertainty quantification methods that are both mathematically rigorous and computationally efficient remains an important open challenge for the field \cite{Simm2017, Reiher2022, Grasselli2025}.

To ensure that quantum chemical calculations are no longer necessary, broad, domain-specific studies are required to show in which areas MLIPs are reliable. Then, MLIPs can be used without resorting to expensive quantum mechanical methods, similar to how DFT is currently used without validating calculations with more accurate methods.

\section{Conclusion and Outlook}
We are midst in a process that will fundamentally change chemistry as a scientific endeavor \cite{Aspuru2018}. Successes in machine learning require us to question
how we do chemical research, what tools we use, and how and what we teach in chemistry \cite{Vogiatzis2026}. 
The efficiency and surprises of machine learning and artificial intelligence approaches affected all branches of chemistry, also how we model chemical processes.

While theoretical chemistry has been based on the modeling of the explicit dynamics of atomistic entities (electrons, atomic nuclei, atoms, united atoms, ...) deeply rooted in mechanical theories well-established in physics, we now see a shift toward implicit models with surprising extrapolation capabilities emerging from data-driven approaches in the field now known as digital chemistry. Thomas Kuhn's concept of a (fundamental) 'paradigm shift' --- unfortunately too often used in an inflationary manner --- is actually seen here in action.

For the construction of potential energy surfaces, routine physics based modeling will be replaced by implicit models that, in some impenetrable way, have inferred the physics from large data sets. Foundation MLIPs will become the starting point of computational chemistry campaigns, completely replacing DFT for this purpose. While their performance is already very impressive, future work along the lines of the open challenges discussed in this paper will further increase their reliability, accuracy, and general applicability. Accurate ab initio reference data will replace the DFT training data so that they can achieve (multi-reference) coupled cluster accuracy from the outset, beating DFT methods not only in terms of computational costs, but also in terms of accuracy. Moreover, specific fine-tuning will always be possible thanks to rolling benchmarking calculations which can produce system-focused ab initio data in an automated way. Accordingly, accurate quantum chemical methods (obtained by traditional or quantum computation) will be used for generating training and validation data in routine computational chemistry. 

While quantum chemical approaches have always come with the promise of rigor and systematic improvability (a desire, too often torpedoed by the curse of dimensionality in quantum mechanics), accurate error bars have remained unaccessible for some specific problem at hand in routine applications \cite{Reiher2022}. By contrast, machine learning approaches usually come with built-in uncertainty quantification, which, if reliable, allow for system-focused studies that no longer require a well-defined underlying model, but a reliable uncertainty measure. This allows for a mix and match of machine learning interatomic potentials, which is important for their system-specific fine tuning. Instead of using acronyms for approximations (as in DFT), we will eventually only resort to error bars obtained from faithful uncertainty quantification procedures.

When the next-generation foundation MLIPs with automated fine-tuning and uncertainty quantifcation becomes available in standard software for computational chemistry, they will be tested by the community and they will convince the community because of their easy applicability, accuracy, and speed and because of the system sizes or number of systems that become computationally accessible. Akin the DFT revolution in computational chemistry, we are now at the brink of the MLIP revolution where MLIP applications will become the new canonical approach to modelling chemistry. If the software implementations of these advanced MLIPs become sufficiently user friendly and black box, one might not even note the shift from density functionals to MLIPs, except that the calculations are now more accurate and faster.

Accordingly, we may expect a whole new jargon in computational chemistry. But it will not only be the way we speak about computational chemistry simulations in routine applications that will change, we can even anticipate a whole new class of chemical concepts that will allow us to extract descriptive and predictive qualitative information out of MLIP calculations, very much like standard chemical reactivity concepts that were rooted in quantum chemistry (especially within conceptual DFT).

MLIPs are also the data-driven solution to unite the fields of quantum chemistry and materials science (electronic structure) with
polymer and (bio)molecular simulations (force-field based configurational screening): quantum accuracy of microstate energies will be so quickly accessible that enhanced sampling techniques can be exploited to represent configurational flexibility properly, but with quantum accuracy, a key to understanding all sorts of complex and functional processes in the molecular sciences.

\section{Acknowledgment}
This work was created as part of NCCR Catalysis (grant number 180544), a
National Centre of Competence in Research funded by the Swiss National Science Foundation.

\providecommand{\latin}[1]{#1}
\makeatletter
\providecommand{\doi}
  {\begingroup\let\do\@makeother\dospecials
  \catcode`\{=1 \catcode`\}=2 \doi@aux}
\providecommand{\doi@aux}[1]{\endgroup\texttt{#1}}
\makeatother
\providecommand*\mcitethebibliography{\thebibliography}
\csname @ifundefined\endcsname{endmcitethebibliography}
  {\let\endmcitethebibliography\endthebibliography}{}

\end{document}